\documentclass[11pt ]{article}

 \usepackage{color}    
 \usepackage{graphicx} 
\usepackage{amsmath}
\begin{document}

\title {Limit  on Continuous Neutrino Emission from Neutron Stars }
\author{Itzhak Goldman$^{1,}$\footnote{Itzhak Goldman thanks the Department of Astronomy and Astrophysics, Tel Aviv University for
the hospitality while on Sabbatical Leave from Afeka College.}  and  Shmuel Nussinov$^{2, 3}$ } 
  \maketitle
  $^1$Department of Exact Sciences, Afeka College of Engineering, Bnei Ephraim 218, Tel Aviv 69107, Israel, email: goldman@afeka.ac.il \\
\\ $^2$School of Physics and Astronomy, Tel Aviv University, Ramat Aviv, Tel Aviv 69978, Israel, email: nussinov@post.tau.ac.il \\
\\ $^3$Schmid College of Science,Chapman University, Orange, California 92866, USA

\begin{abstract} 
 The timing data of  the binary pulsar   PSR1913+16, are used to establish an upper limit on the rate of continuous neutrino emission from neutron stars. Neutrino emission from each of the neutron stars  of the binary system, increases the star binding energy and thus
translates  to a decrease in their masses. This in turn implies an increase with time   of the binary period. Using the pulsar data  we obtain an   upper limit  on the allowed rate of mass reduction :
$| \dot{M}| <1.1 \times 10^{-12} yr^{-1} M $, where  $M$  is the total mass of the binary. This constrains exotic
nuclear equations of state that predict continuous neutrino emissions. The limit applies also to other channels of energy loss, e.g. axion emission. Continued timing measurements of additional binary pulsars, should yield a stronger limit in the future.  
\end{abstract}

 keywords: neutrinos - neutron stars - nuclear matter - pulsars

\section{Introduction}

Neutron stars form via the core collapse of massive stars. Most of the
 gravitational energy ( $\sim 0.15 M_{core}c^2\sim 3\times 10^{53}$ erg ) is emitted during the
first $O(10) $ seconds via $O(10-20) MeV$ neutrinos, from the neutrino-sphere of the proto neutron star, of temperature
 $T \sim$  few MeV. This old prediction was experimentally verified in the neutrino
pulses from SN 1987A observed at IMB and Kamiokande underground detectors. Neutrino energy losses in stellar interiors are also relevant  in the pre-supernova stage\cite{Itoh}.  

 Further cooling of the neutron star via neutrino emission is likely to continue
 for a long time. A famous cooling mechanism is the URCA two-step process:

$$  n\to p+e^-+\bar{\nu_e} ;\ \ \  e^-+p \to n +\nu_e$$

 or its modified form:

$$ n+n \to n + p + e^- + \bar {\nu}_e; \ \ \ n+p +e^-  \to n+n + \nu_e$$
  After about $10^5$ years the combined emission by neutrinos and photons drops below $10^{35} erg s^{-1}$\cite{page}.
 
Recent nuclear experiments   \cite{subedi}, \cite{piasetsky}  have found strong neutron-proton correlations. This in turn can motivate non-standard nuclear-matter models \cite{frankfurt}  which, in particular, may  imply that neutron stars  emit neutrinos {\it continuously}. 

In the present work we wish to set observational limits on{\it any model}   that predicts {\it a continuous}
emission of neutrinos from neutron stars.

The continuous emission of neutrinos increases the star binding energy and thus reduces   the mass of the neutron star. If the neutron star is a member of a stellar binary system, such a reduction will cause, as detailed below, an increase with time of the stellar binary period. In this short paper we   employ the timing data of the binary pulsar PSR1913 +16 to set an upper limit on the rate of mass reduction in this system. 
For a given  energy spectrum this limits the neutrino  number luminosity.

\section{Binary orbital period change  due to neutrino emission}

  Jeans (1924) \cite{Jeans}, pointed out that   the mass of a star which emits electromagnetic radiation decreases with time and therefore 
the orbital elements of a binary system should evolve with time. Assuming that in the local frame of each star the  
emission is  spherically symmetric, he obtained

\begin{equation}
M a =\ {\rm constant}
\end{equation} 

where $a$ is the semi-major axis and $M=m_1 + m_2$ is the total mass of the system.
This and the  expression for the binary period

\begin{equation}
P_b = 2\pi\sqrt{\frac{a^3}{G M}}
\end{equation}

imply

\begin{equation}
\frac{\dot{P}_b}{P_b}=  -2 \frac{ \dot M}{M}
\end{equation}  
  Since $ \dot M <0 $,  $ \dot{P}_b >0 $ so that  the orbital period keeps increasing.
   
Except for the nature of emission process this is the same as the question addressed here.
 
We wish to use the limits on the residual rate of change of the orbital period of the binary pulsar PSR1913+16, after accounting for the rate due to emission of gravitational radiation, to  limit  the rate of mass decrease due to neutrino emission.
 
  The pulse period $P_p= 59ms$ and the pulse period derivative $\dot P_p= 8.63 \times 10^{-18}$ of PSR1913+16 \cite{wt}, \cite{will} imply a spin down age of $4.34\times 10^8$ years.  The orbital period timing data  are listed below. The first two are the observed binary period, and the observed time-derivative of the binary period,  corrected for effects of the acceleration due to the Galaxy. The third is the predicted decrease of the binary period due to emission of gravitational radiation.
 
\begin{equation}
P_b= 0.322997 Day = 27907 s 
 \end{equation}
\begin{equation}
 \begin{split}
 &\dot{P}_{b, Corrected} =  \dot{P}_{b, Obs} - \dot{P}_{b, Galaxy}= \\
& {} - (2.4056 \pm 0.0051)  \times 10^{-12} 
   \end{split}
 \end{equation}
 \begin{equation}
\dot{P}_{b, GR} =-(2.40242  \pm  0.00002) \times 10^{-12}
\end{equation}
 so that the unaccounted for time-derivative of the binary period is
 \begin{equation}
 \begin{split}
&\dot{P}_{b, residual}=\dot{P}_{b, Corrected} - \dot{P}_{b, GR} \\ &= (  - 3.2 \pm  5.1 ) \times 10^{-15}
\end{split}
\end{equation}    
  
 yielding
 \begin{equation}
\frac{1}{P_b}\dot{P}_{b, residual} =( - 3.6 \pm  5.8) \times 10^{-12} yr^{-1}
\end{equation}     
 By attributing all  of the positive  residual rate to neutrino emission and using Eq. (3), it is possible to obtain  a conservative limit on the rate of  mass loss due to   neutrino emission.
 
  \begin{equation}
  \frac{| \dot M|}{M} \leq 1.1  \times 10^{-12} yr^{-1}
\end{equation} 
   
    \section{The surface temperature and electromagnetic luminosity}

In the following  we discuss the electromagnetic signature of   the energy generated by instabilities arising from non standard nuclear matter equations of state. The total generated energy corresponding to the limit of Eq.(9) is

 \begin{equation}
  L^*=\dot M  c^2 = \frac{M}{M_{\odot}} 6.29\times 10^{34} erg/{sec}  
\end{equation}         
If a non negligible fraction of   this  luminosity would reach the surface of the neutron star, rather than escape as volume emission of neutrinos, the surface temperature
and electromagnetic luminosities of isolated neutron stars would be  significantly  higher than observed.
This in turn would have allowed  a tightening of the constraint   of Eq.(9) .

We follow Yakovlev \& Pethick (2004)\cite{yakpeth} and examine the energy transport in the core of the  neutron star.
 To this end we solve first  the  Tolman-Oppenheimer-Volkoff\cite{ST} general relativistic structure equations for two nuclear matter equations of state derived recently from observed 
masses and radii of neutron stars by Steiner, Lattimer \& Brown\cite{steiner}.

The  static spherically symmetric   line-element squared can be written as

\begin{equation}
ds^2= e ^{2 \phi(r)}c^2 dt^2 - e^{2\lambda(r)}dr^2 - r^2  d\Omega^2
\end{equation}   
  
With $\epsilon(r)$ and $p(r)$ the energy density and the pressure   respectively, the structure equations are:
 \begin{eqnarray}
m'(r)= 4\pi r^2 \epsilon(r)\\
e^{2\lambda(r)} =\left(1- \frac{2 G m(r)}{c^2 r}\right)^{-1}\\
p'(r)= -\frac{G M  e^{2\lambda(r)}}{ c^2 r^2}\left(\epsilon(r) + p(r)\right)\left(1+ 4 \pi \frac{ p r^3}{m(r) c^2)}\right)\\
\phi '(r)= -\frac{p'(r)}{\epsilon(r) + p(r)}
 \end{eqnarray}
 The mass density   $ \rho(r) = m_p n(r)$, where  $n(r)$ is the nucleon number density is obtained from
 
 \begin{equation}
\frac{ \rho '(r)}{\rho(r)}= \frac{\epsilon '(r)}{\epsilon(r) + p(r)}
\end{equation}
  where a prime denotes radial derivative.
  
  The nuclear equations of state employed here are  those presented in Tables 5 and    6 of \cite{steiner}. 
  The first one is softer than the second. Together, they bracket the equations of state that can reproduce the observational data considered by \cite{steiner}. For each of the above  we have computed two models with different value of the central mass density.   For each model we obtain the r-dependence of  the metric,  
  pressure, and  mass   and energy densities.
  
  The very high thermal conductivity of the core implies \cite{yakpeth} that the redshifted temperature
$T(r) e^{\phi(r)}$ is constant throughout the core so the temperature at any radius is determined by the central temperature $T_c=T(0)$:

\begin{equation}
T(r) = T_c e^{\phi(0) -\phi(r)}
\end{equation}

In a steady state the sum of neutrino emission  from the core (volume integrated) and the electromagnetic luminosity from the stellar surface should  equal $L^*$. 

\begin{equation}
\int_0^R 4 \pi r^2 Q_{\nu}(T(r)) e^{2\phi(r)} dr + 4 \pi R^2 \sigma T_s^4 e^{2\phi(R)}      =L^*
\end{equation}
where $Q_{\nu}$ is the neutrino volume emissivity and $T_s$ is the surface temperature.
 
We follow the prescription of \cite{yakpeth} that in high density regions the direct (i.e. original) URCA process is allowed in addition to the modified URCA process,  whereas in  lower density regions only the latter Is allowed.
Thus we apply

\begin{eqnarray}
Q_{\nu}(T) = \left(1.5\times 10^{15} T_7^6 + 2.48\times 10^{5} T_7^8\right) erg \ cm^{-3} \ ,  \ \ \rho> 7.85\times 10^{14}gr cm^{-3}\\
Q_{\nu}(T) = 2.48\times 10^{5} T_7^8  erg \ cm^{-3} \ ,\  \ \rho< 7.85\times 10^{14}gr cm^{-3} 
\end{eqnarray}
where $T_7 = T/(10^7 K)$
 For   weak magnetic fields there exists a relation between the surface temperature and the core temperature at its boundary $T_b$ with the outer crust. The relation derived by Gudmundsson et al. (1983)\cite{Gudmund} and quoted by \cite{yakpeth} is

\begin{equation}
T_b =   1.288\times 10^8 \left(\frac{T_{s6}}{g_{14}}\right)^{0.455}K \ , \  \  g_{14}= \frac{G M e^{-\phi(R)}}{R^2 10^{14} cm\  s^{-2}} \ , \ \ T_{s6}=\frac{T_s}{10^6 K}
 \end{equation}

For a given neutron star model    $T_c$ is varied until Eq. (18) is satisfied. This in turn determines the values of 
 the surface temperature and electromagnetic luminosity.
 The results are presented in Table 1. Here, $L_{\infty}$ and $T_{s, \infty}$ are the electromagnetic luminosity and surface temperature observed at infinity.

In the first three models the central density is high enough so that the direct URCA model is operational in the inner part of the core and the modified URCA process takes over in the outer part of the core. In the fourth model the central density is too low to allow the direct URCA process and only the modified process is operational throughout the core.

For the first models the electromagnetic luminosity is about $2.5\times 10^{-4}$ of the total luminosity while for the fourth it is  $\sim \ 3\times 10^{-2}$ of the total luminosity. In all cases, the ensuing surface temperature and   electromagnetic luminosity are at the lower end of the corresponding observed values\cite{yakpeth}.
Therefore, this would not be a useful way to constrain the non standard equations of state.  
 
\begin{table}[h]
\caption{Neutron star models }
     \vskip 0.5 truecm 
\begin{tabular}{ p{0.4 cm}     p{2cm}  p{1.8cm} p{1.8cm}p{1.8cm}p{1.8cm} p{1.8cm}p{1.8cm}} 
\hline 
  
 Eos  & $ \rho_c [10^{14} gr  cm^{-3} ]$  &\  \ $M[M_{\odot}]$  & R [km ]&   ${\hskip -0.5 truecm}L^* [10^{34} erg\ s^{-1} $]   &  $ T_c [10^7\ K]$ & ${\hskip -0.3 truecm}T_{s,\infty} [10^5\ K]$  & ${\hskip -0.3 truecm}L_{\infty}[ 10^{31} erg\ s^{-1}]$ 
 \\
  \hline  
  1& 11.5  &  1.55  &  12.05   & 11.88  &  2.33 &  3.77 &1.28  
 \\
\hline
  1& 10.7  &  1.45  &  11.94  &  9.11  &  2.38 &  3.73 &1.27 
   \\
  \hline  
2& 8.5  &  1.78 &  12.44   & 11.85 &  2.85 &  4.55 &2.72  
 \\
\hline
  2& 7.47  &  1.55  &  12.2   & 9.74  &  25 & 14.7 &320
   \\
   \hline
   \end{tabular} 
 {\footnote \ Eos given in Table 5 of \cite{steiner}\ \ }
 {\footnote \ Eos given in Table 6 of \cite{steiner}}
\end{table}

\section{Discussion}
This paper has been motivated  by the observation of Leonid Frankfurt that the
strong n-p pairing correlations at short distances discovered at Jefferson Lab
\cite{subedi} can significantly modify the nuclear equation of state and consequently, the physics of neutron stars.  In particular it is conceivable that these eventually lead to an instability of
neutron stars. 
However, the bound derived here is a {\it theory-independent} constraint that must be 
satisfied by {\it any} non-standard nuclear-matter model which predicts {\it continuous} neutrino emission from neutron stars. 

The neutrino luminosity corresponding to the limit obtained here is about $ 10^{35} erg s^{-1}$ a value that exceeds the combined neutrino and photon luminosities for neutron stars older than about $10^5$ years \cite{page}.

The rate of energy loss from PSR1913+16 due to emission of electromagnetic radiation and particles by the pulsar can be estimated by use of the observed slow down of the pulsar\cite{SB}. Using the updated observational values this rate is  $\leq 1.66\times 10^{33} erg s^{-1}$ about two orders of magnitude smaller than the limit of Eq. (9).

We note that the limit applies also to scenarios in which the energy emission is by axions.  Continued timing measurements of additional binary pulsars, notably the double pulsar PSR J0737-3039,  should yield a stronger limit in the future.

We considered the transport of the generated energy through the neutron star, employing the equations of state
obtained by \cite{steiner} on the basis of observations.  We found that the lion share of the luminosity  is due to the neutrino volume emission. For neutron star models with high central density the surface temperature was found to be $T_{s,\infty}= (3.73 - 4.55)\times 10^5 K$.
For a less dense neutron star  $ T_{s,\infty}= 1.47 \times 10^6 K$.   These values are at the lower end of the   observed surface temperatures\cite{yakpeth}.
As a result, the surface temperature and the related electromagnetic luminosity are low enough so that they would not serve as means to constrain  the non standard equation of state.

It was suggested to us by Clifford Will \cite{cwill} that timing data of pulse period derivative could be useful in  constraining  the mass loss.   Indeed, for some millisecond pulsars  values of $\frac{\dot P_p}{P_P}= \sim 2 \times 10^{-11} yr^{-1}$  after correction for the proper motion effect have been included\cite{Kizil}. Moreover, the uncertainty is about 5 orders of magnitude smaller.
The mass loss will induce a change in the moment of inertia and thus  a change in the period in addition to the  one attributed to energy loss by a rotating magnetic dipole.
However, in contrast with orbital period decay where there is a robust estimate of the value of the 
period change due to gravitational radiation, here there is no independent measurement for the magnetic field.
Thus, the high precision is not useful and only the total value of the rate of change of the pulse period can be used. This is more than an order of magnitude weaker than our limit Eq. (9).

 In spite the relatively large neutrino luminosity corresponding to the limit obtained here, its  contribution to the   background diffuse integrated neutrino luminosity   is negligible compared to that  due to all
cosmological supernovae \cite{HBD} events. Moreover, the much lower neutrino energies render the detection much more difficult. 

\subsubsection*{Acknowledgments}
We have greatly benefited from many discussions with L. Frankfurt and E.Piasetsky. We thank C. Will for 
bringing up  pulsar spin-down. We thank the referee for his  useful   comments.

\end{document}